# An Algorithm for Constructing All Families of Codes of Arbitrary Requirement in an OCDMA System

Xiang Lu, Jiajia Chen and Sailing He, *Senior Member, IEEE*

*Abstract*—A novel code construction algorithm is presented to find all the possible code families for code reconfiguration in an OCDMA system. The algorithm is developed through searching all the complete subgraphs of a constructed graph. The proposed algorithm is flexible and practical for constructing optical orthogonal codes (OOCs) of arbitrary requirement. Simulation results show that one should choose an appropriate code length in order to obtain sufficient number of code families for code reconfiguration with reasonable cost.

## I. INTRODUCTION

OPTICAL code division multiple access (OCDMA) is a relatively new bandwidth-shared method for various kinds of network. Compared with traditional bandwidth-shared methods such as time division multiplexing (TDM) and wavelength division multiplexing (WDM) systems, OCDMA has many advantages such as high security and capacity potential [1]-[4]. In an OCDMA network each user can only get the information associated with him through his own decoding key.

In an OCDMA system it is important to select suitable codes for all the online users, and the system performance, such as the maximum number of the users that can be supported and the bit error rate (BER), depends on the codes selected. The security of the network also depends on the available code space. Usually optical orthogonal codes (OOC) perform well due to their low correlation value and relatively large code space. Typically an OOC can be described by a set of parameters $(n, \omega, \lambda_a, \lambda_c)$, where $n$ is the length of the code, $\omega$ denotes the weight of the code which determines the total number of '1' in the code, $\lambda_a$ is the auto-correlation limit, and $\lambda_c$ is the cross-correlation limit. In order to improve the security and eliminate the vulnerability to eavesdropping, a strategy of reconfigurable (or switchable) codes is preferred (see e.g. [1, 4]). Therefore, previous works on constructing a family of optimal $(n, \omega, \lambda_a, \lambda_c)$ OOC codes (see e.g. [5, 6]) are not suitable for code reconfiguration. In this paper, we present a simple construction algorithm, which can guarantee to find all the families of the codes under an arbitrary constraint, for the purpose of family of reconfigurable OOC

## II. THE PRESENT CODE CONSTRUCTION ALGORITHM

We also use a parameter $C$ to determine the capacity of the code family (i.e., how many codes can be supported in the system). Obviously a large $C$ means the system can support more users at the same time. However, C is limited by parameters $n$, $\omega$, $\lambda_a$, and $\lambda_c$ (usually $\lambda_a = \lambda_c = \lambda$). This limit is called Johnson limit and is give by [5, 6]

$$C = \left\lfloor \frac{(n-1)(n-2)...(n-\lambda)}{\omega(\omega-1)(\omega-2)...(\omega-\lambda)} \right\rfloor \qquad (1)$$

To construct a code family $(n, \omega, \lambda_a, \lambda_c)$, first we introduce some definitions:

1. Distance (between two '1's): the number of chips between two '1's in a code. For example, the distance of two '1's in code '1000100000' is 4 (the code should be viewed as chips in cycle, i.e., the last chip should be considered as the adjacent chip to the first one). Note that the number of chips can be counted clockwise (in one direction) from the first chip to the second chip or from the second chip to the first chip, and we need to choose the smaller value as the distance (e.g., the distance of the two '1's in code '100010' is 2). We also use a parameter $d = \lfloor n/2 \rfloor$ to denote the diameter of a code with length n. Obviously the distance must be no larger than $d$.

2. Distance vector: Distance vector is a vector consisting of many elements which give the inter-distances of several '1's. In the 1-distance vector, the element is the distances of any two

Manuscript received January 18, 2006. This work was partially supported by Natural Science Foundation of Zhejiang Province of China (under grant No. R104154)

All the authors are with Centre for Optical and Electromagnetic Research, Joint Laboratory of Optical Communications of Zhejiang University, Zhejiang University, Hangzhou 310058, China.

X.Lu is also with Laboratory of Optics, Photonics and Quantum Electronics, Department of Microelectronics and Information Technology, Royal Institute of Technology (KTH), Electrum 229, 164 40 Kista, Sweden.

S. He is also with Division of Electromagnetic Engineering, School of Electrical Engineering, Royal Institute of Technology (KTH), 100 44 Stockholm, Sweden.

Address correspondence to Sailing He. E-mail: sailing@kth.se.

'1's in a code. For example, the 1-distance vector of code '1000100000100000000' is $[4,6,9]$. Since in an OCDMA system the rotation of a code represents the same code, $[4,6,9]$ represents all the rotations of code '1000100000100000000', such as code '0100010000010000000', etc. In fact, the above definition is for 1-distance vector, which can be generalized to a $\lambda$-distance vector.

We construct a $\lambda$-distance vector for each found code as follows. We first find all its subcodes (of the same length n) with ($\lambda$+1) number of '1's (and the rest chips set to 0) and put them in an order $\left[ subcode^1, subcode^2, ..., subcode^S \right]$, where $S = C_\omega^{\lambda+1}$. Then we find the 1-distance vector $(d_1^s, d_2^s, ...)$ for the s-th subcode. The $\lambda$-distance vector is formed by $\left[ (d_1^1, d_2^1, ...), (d_1^2, d_2^2, ...), ..., (d_1^s, d_2^s, ...) \right]$. For example, the 2-distance vector of code 1101001000000 can be expressed as $\left[ (1,2,3), (1,5,6), (3,3,6), (2,3,5) \right]$

If the auto-correlation limit of a code is 1, all the elements of its 1-distance vector must be different. Otherwise, a code after some rotation can overlap with the original code by two '1'-chips (i.e., with the same distance) and consequently violate the requirement of auto-correlation limit of 1. For the same reason no element of the 1-distance vectors for two different codes should be the same if the cross-correlation limit is 1. In the general case the requirement of correlation limit $\lambda$ is equivalent to the following requirement for a code family: each two codes should not have any same element in their $\lambda$-distance vectors.

Here we describe the presented algorithm with an example of constructing all possible code families of requirement $(19,3,1,1)$. Obviously a 1-distance vector of each code (with diameter d) can have $d$ elements at the most. Code weight 3 means the total number of elements in the 1-distance vector is 3 for each code. The available values for these elements are integers from 1 to 9 in the case of $n = 19$. Thus, we can obtain $C = 3$ codes at the most for the requirement of $(19,3,1,1)$.

First we search all the possible codes satisfying $n = 19$, $\omega = 3$, and $\lambda_a = 1$. Since a 1-distance vector is determined from an actual code, the elements in a 1-distance vector cannot be arbitrary and must satisfy some constraint. For example, 1-distance vector [3, 3, 7] dose not exist for $n = 19$. The constraint can be expressed as follows:

$$x_1 + x_2 + x_3 = 19 \qquad (2)$$

where $x_i \left( when\ x_i \leq 9 \right)$ or $19 - x_i \left( when\ x_i > 9 \right)$ represents a distance in the 1-distance vector of a code. In other words, each vector satisfying condition (2) represents a 1-distance vector. For example, $\left[ x_1 = 4, x_2 = 6, x_3 = 9 \right]$ represents a 1-distance vector [4,6,9], and [x₁=2, x₂=3, x₃=14] represents a 1-distance vector [2,3,5]. All the 1-distance vectors can be obtained by solving (2) in the case of $n = 19$.

Then we search a family of codes satisfying $\lambda_c = 1$ from all the 1-distance vectors found above. From the Johnson limit we know the capacity of this family is $C = 3$. We can find a suitable family consisting of 3 orthogonal codes through the help of some graph theory. Consider each solution (such as 1-distance vector $[4,6,9]$) to Eq (1) as a node in a graph. If any two nodes of this graph satisfy the requirement of $(19,3,1,1)$, we make an edge connecting these two nodes. If two 1-distance vectors (corresponding to two codes) have a same element, these two codes cannot be put in the same code family and cannot be connected with an edge. The final graph $G(V, E)$ tells the relationship of the coexistence of the codes obtained. The complete graph with c nodes connecting to each other (called c-complete graph) determines the c-code families which satisfy the requirement of $(19,3,1,1)$. Here $c = C, C-1, ..., 1$. In other words, the problem of searching code families with a specific number of orthogonal codes can be seen as a problem of searching completed sub-graphs with a specific number of nodes in graph G. Such a searching for completed sub-graphs can be easily carried out with a numerical simulation (see Fig. 1 for a pseudo-code).

```
find c - complete subgraph{
    for(i = 1; ∃c nodes in graph G(V, E); i + +){
        find all the (c - 1) - complete subgraphs in a
            graph formed by nodes connecting to nodeᵢ;
        record each (c - 1) - complete subgraph and
            nodeᵢ as a c - complete subgraph;
        delete nodeᵢ and its corresponding edges from G(V, E);
}}
```

Fig. 1 Pseudo-code for searching completed sub-graphs with c nodes in graph.

III. GENERALIZATION OF THE ALGORITHM

After illustrating the present algorithm with the above example, we generalize the algorithm in the following steps for searching all the possible families of OOCs with the general requirement $(n, \omega, \lambda_a, \lambda_c)$.

Step 1:
Solve the following equation of constrain:
$$x_1 + x_2 + ... x_\omega = n$$
$$x_i > 0, i = 1, 2, ..., \omega. \quad (3)$$

Step 2:
If $x_i$ is larger than $d$, replace the original $x_i$ with ($n - x_i$).

Step 3:
For auto-correlation limit $\lambda_a$, we use $\lambda_a$-distance vectors to search the codes. If all the elements of a $\lambda_a$-distance vector are different from each other, the auto-correlation limit of the corresponding code is no larger than $\lambda_a$, and thus we can save such a code. We construct a graph G(V,E), and each saved code can be seen as a node in this graph.

Step 4:
Consider two $\lambda_c$- distance vectors of found codes. If the elements of these 2 vectors are totally different, we can say the corresponding two codes can coexist in the same code family. Then the corresponding two nodes can be connected with an edge in our constructed graph G(V,E).

Step 5:
After constructing the graph G(V,E). We need to find a c-completed subgraph. Here c is the required total number of the codes in our code family. If c equals to the Johnson's limit then the code family we find is optimal. The algorithm of searching the complete subgraph is the same as that for the case of 1-distacne vectors (see Fig.1).

### IV. SIMULATION RESULTS

The algorithm is run with a Visual C++ code in a PC, Fig. 2 shows the total available number of some specific OOC families (with c number of codes in a family; $\omega$ =3, and $\lambda_a = \lambda_c = 1$) as the code length increases.

In this figure the code length n is represented with the positive offset value (i.e, $n - n_J$) to the optimal Johnson number $n_J$. The optimal Johnson number is the minimal code length to ensure the existence of a family with C number of codes, and is given by (when $\lambda = 1$)
$$n_J = C \cdot \omega \cdot (\omega - 1) + 1 \quad (4)$$

For example, in (19, 3, 1, 1), $n_J$ =19 for a family of 3 codes because we cannot get any family of 3 codes if the code length is less than 19. From this figure one sees the total number of available code families increases quickly as the offset increases.

This indicates that in the construction of the codes with specific $(\omega, \lambda_a, \lambda_c)$ (note that n is flexible here) a large offset (i.e., a larger n) can give more families of available codes and consequently is more advantageous for the strategy of code reconfiguration. This is valuable in a network requiring a high security. From the simulation results one can also see that only a small offset is needed (e.g. the total number of available code families increases by at least 1 order when n increases from $n_J$ to $n_J + 1$) in order to get sufficient number of code families (a large offset will increase the cost of the OCDMA system). From this figure one sees that for a fixed length of codes, a larger number of codes in a family will decrease the total number of available code families. For example, for the case of n= 25, c=4 gives over 200 code families while c=3 (n= 25 requires offset = 6) gives over 2000 code families. In other words, for the strategy of code reconfiguration, one should choose a suitable value of c (a bit smaller than the Johnson limit C) for a family.

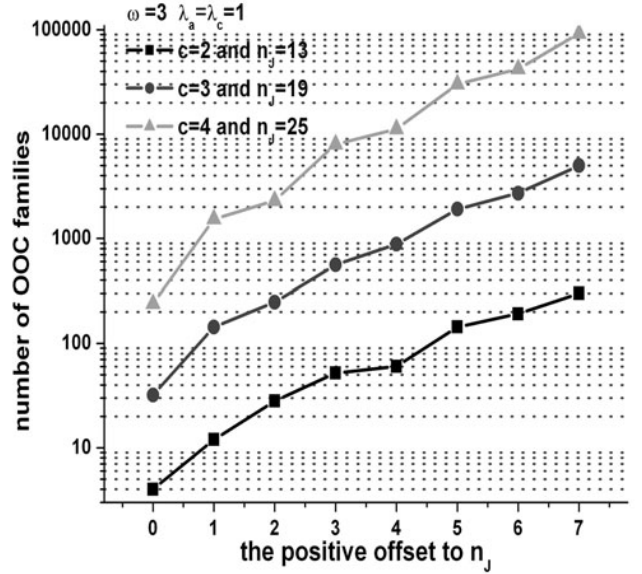

Fig. 2 The total number of OOC families as the positive offset value to $n_J$ increases

### V. CONCLUSION

In this paper we have presented a novel OOC code construction algorithm which can find all the possible code families with an acceptable speed. Unlike the other code construction algorithms, the present algorithm is flexible and practicable for constructing OOCs of arbitrary requirement, and can guarantee to find all the families of required OOCs. With the present algorithm, one can choose an appropriate value c according to the actual requirement for the total number of the codes in a family in order to obtain sufficient number of OOC families for code reconfiguration. Our numerical simulation results have also shown that one only needs to use a small offset to the optimal Johnson number $n_J$ in order to get sufficient number of code families.